\documentclass[a4paper,12pt]{article}
\usepackage{amssymb}
\usepackage{float}
\usepackage{times}
\usepackage{graphicx}
\usepackage{amsmath}
\usepackage{subfig}
\usepackage{bm}
\usepackage{color}
\usepackage{natbib}

\newtheorem{Thm}{Theorem}

\newtheorem{Lemma}[Thm]{Lemma}
\newtheorem{Cor}[Thm]{Corollary}

\begin{document}
\thispagestyle{empty}

\begin{center} 
{\Huge\bf A new distribution function with bounded support: the reflected Generalized Topp-Leone Power Series distribution}\\
\ \\ \ \\ \ \\ \ \\
{\large\bf Francesca Condino} \hspace{1 cm} \textbf{and} \hspace{1 cm} {\large\bf Filippo Domma *} \\ \vspace{1 cm}
{
Department of Economics, Statistics and Finance \\
University of Calabria  - Italy.  \\
\textbf{f.domma@unical.it} \hspace{0.3 cm}  and \hspace{0.3 cm}   \textbf{francesca.condino@unical.it}\\ [1cm] }
\end{center}
\vspace{5 cm}
{\large\bf * Corresponding author: f.domma@unical.it }

\newpage
\setcounter{page}{1}
\topmargin 2 cm
\begin{center}
{\Huge\bf A new distribution function with bounded support: the reflected Generalized Topp-Leone Power Series distribution}  \\ \ \\
\ \\ \ \\

\begin{abstract}
\textit{In this paper we introduce a new flexible class of distributions with bounded support, called reflected Generalized Topp-Leone Power Series (\textit{rGTL-PS}), obtained by compounding the reflected Generalized Topp-Leone \citep{vanDrop06} and the family of Power Series distributions. The proposed class includes, as special cases, some new distributions with limited support such as the \textit{rGTL}-Logarithmic, the \textit{rGTL}-Geometric, the \textit{rGTL}-Poisson and \textit{rGTL}-Binomial.
This work is an attempt to partially fill a gap regarding the presence, in the literature, of continuous distributions with bounded support, which instead appear to be very useful in many real contexts, included the reliability. Some properties of the class, including moments, hazard rate and quantile are investigated. Moreover, the maximum likelihood estimators of the parameters are examined and the observed Fisher information matrix provided. Finally, in order to show the usefulness of the new class, some applications to real data are reported.}

\end{abstract}
\ \\
\end{center}
\vspace{1  cm} {\bf Key words}: Compound Class, Bounded Support, Hazard Function, Flexible shape.  \\

\baselineskip 0.7 cm

\section{Introduction}
In recent years, many authors have focused their attention on the proposition of new and more flexible distribution functions, constructed using various transformation techniques such as, for example, the Beta-generated distribution by \citet{Eugene02} and \citet{Jones04}, the Gamma-generated distribution by \citet{Zagrafos09}, Kumaraswamy-generated distribution by \citet{Cordeiro11}, McDonald-generated distribution by \citet{Alexander12}, \citet{Ristic12}, Weibull-generated distribution by \citet{Bourguignon14}, just to name a few. Such transformation techniques may be viewed as special cases of the Transformed-Transformer method proposed by \citet{Alzaatreh13}. Other proposals are based on the Azzalini's method \citep{Azzalini85, Azzalini86} and its extensions \citep{Domma15}. Using these techniques, many of the known distributions such as, for example, Normal, Exponential, Weibull, Logistics, Pareto, Dagum, Singh-Maddala, etc., have been generalized. \\
Almost all these new proposals concern distributions with unbounded support. In the face of the numerous proposals of distributions with unbounded support emerges, undoubtedly, the great scarcity of distributions with bounded support \citep[pag. 473]{Marshall07}, although there are many real-life situations in which the observations clearly can take values only in a limited range, such as percentages, proportions or fractions. \citet{Papke96} claim that variables bounded between zero and one arise naturally in many economic setting; \textit{e.g.} the fraction of total weekly hours spent working, the proportion of income spent on non-durable consumption, pension plan participation rates, industry market shares, television rating, fraction of land area allocate to agriculture, etc.  Various examples of proportions in the unit interval used in empirical finance are discussed in \citet{Cook08}.
\\
Also in reliability analysis, different authors refer to continuous models with finite support in order to describe lifetime data. This is often motivated by considering physical reasons such as the finite lifetime of a component or the bounded signals occurring in industrial systems \citep[see, for example,][]{Jiang13, Dedecius13}. In this perspective, the models with infinite support can be viewed as an approximation of the realty.
Furthermore, when the realiability is measured as percentage or ratio, it is important to have models defined on the unit interval \citep{Genc13} in order to have plausible results.\\

It is well known that the most used distribution to model continuous variables in the unit interval is the Beta distribution. The popularity of this distribution is certainly due to the great flexibility of its density function, in fact it can take different forms such as constant, increasing, decreasing, unimodal and uniantimodal depending on the values of its parameters. This distribution has been used in various fields of science such as, for example, biology, ecology, engineering, economics, demography, finance, etc. (for a detailed discussion see \citet{Johnson95}, \citet{Nadarajah07}). On the other hand, mathematical difficulties underlying the use of this model, due to the fact that its distribution function cannot be expressed in closed form and its determination involves the incomplete beta function ratio, are well known. \\
Recently, several authors have proposed an alternative to the Beta distribution by recovering the distribution proposed by Kumaraswamy in 1980 in the context of hydrology studies.  As pointed out by \citet{Jones09}, Kumaraswamy's distribution has many of the properties of the Beta distribution and some advantages in terms of tractability, in particular its distribution function has a closed form and it does not involve any special function.  In fact, this distribution turns out to be a special case of the generalized Beta of the first type proposed by \citet{McDonald84} (see \citet{Nadarajah08}).

Perhaps thanks to a work of \citet{ Nadarajah03}, a renewed interest has been recently developed for another distribution defined on bounded support, the Topp-Leone (\textit{TL}) distribution, proposed by \citet{ToppLeone55} and successively studied by different authors \citep[see, for example,][]{Ghitany07, Genc12, Vicari08} also in the reliability context \citep{Ghitany05, Genc13, Condino14}.

Starting from a generalized version of the \textit{TL} distribution, \citet {vanDrop06} proposed the reflected Generalized Topp-Leone (\textit{rGTL}) distribution. Similarly to the Beta distribution, this distribution has a density that can be constant, increasing, decreasing, unimodal and uniantimodal, depending on the values of its parameters. Moreover, it has a strictly positive density value at its lower bound and a closed form of its distribution function. However, the hazard function of the \textit{rGTL} shows a certain rigidity since it is always increasing. This fact is a real weakness of the model, in particular in the field of reliability theory and survival analysis.

With the aim to propose a new flexible model defined on the unit interval, in this paper we consider the standard \textit{rGTL} distribution and  introduce the reflected Generalized Topp-Leone Power Series (\textit{rGTL-PS}) class of distributions, obtained by compounding  the Power Series distributions and the standard \textit{rGTL}.

In the recent literature, many new distributions are obtained by compounding a continuous distribution with a discrete one. An interesting motivation of this procedure can be found in the process underlying the failure of a series system composed by $Z$ component. If $Y_i$ is the lifetime for the $i-th$ component, the system will fail when the first component fails, so the lifetime of the whole system is $Y=\min (Y_1,..., Y_Z)$. In this situation, assuming that the lifetimes of the components are independent, it is easy to obtain the probability of failure for the system by compounding the distribution of $Y$'s and the distribution of $Z$, as follows:
\begin{eqnarray*}
P(Y\leq y)&=&P[\min (Y_1,..., Y_Z) \leq y]=\\
&=&1-\sum_{z=0}^{\infty} P(Y_1>y)P(Y_2>y)...P(Y_n>y)P(Z=z).
\end{eqnarray*}
By considering different cumulative distribution functions for $Z$ and for $Y_i$, various models are proposed during the last years. It is the case, for example, of the exponential-geometric distribution \citep{Adamis98}, of the exponential-Poisson-Lindley distribution \citep{Barreto13} and of the exponential-logarithmic distribution \citep{Tahmasbi08}, to name a few. Other distributions, such as the Weibull Power Series \citep{Morais11}, are obtained by describing the random variable $Z$ through the Power Series distribution, or by considering a similar procedure to that just mentioned, involving the maximum rather than the minimum of the lifetimes \citep{Nadarajah13, Flores13}.

The \textit{rGTL-PS} distribution obtained preserves the main advantage of the \textit{rGTL} distribution with respect to the Beta distribution, that is its cumulative distribution function has a closed form and therefore the quantile functions are easily obtainable and one can easily generate a random variable from \textit{rGTL-PS} distribution. Furthermore, the \textit{rGTL-PS} density function has an analogous flexibility of the Beta density function, \textit{i.e.} the  shape of density can be increasing, decreasing, unimodal and uniantimodal. Finally, the shape of the hazard, besides being increasing and bathtub as the Beta distribution, also shows a N-shape (very useful in the context of reliability theory, see for example, \citet{Bebbington09}, \citet{Lai12}). These properties characterize our proposal as a valid alternative to the Beta distribution.  \\
\\
This article is organized as follows. In \textit{Section 2}, we define the \textit{rGTL-PS} distribution. Some properties, such as the moments, the hazard rate and the quantile are derived in \textit{Section 3}. The maximum likelihood estimation is discussed in \textit{Section 4} and some special cases are studied in \textit{Section 5}. Finally, various applications on real data sets are reported in \textit{Section 6}.

\section{Reflected Generalized Topp-Leone Power Series distribution}
A random variable $Y$ is said to have a standard reflected Generalized Topp-Leone (\textit{rGTL}) distribution  (\cite{vanDrop06}) if its cumulative distribution function (\textit{cdf}) and the corresponding probability density function (\textit{pdf}) are respectively given by
\begin{equation}
G(y;\alpha,\nu)=1-(1-y)^{\nu} \left[\alpha-(\alpha-1) (1-y)\right]^{\nu}
\label{fdr rTL}
\end{equation}
and
\begin{equation}
g(y;\alpha,\nu)=\nu (1- y)^{\nu-1}[\alpha-(\alpha-1)(1-y)]^{\nu-1}[\alpha-2(\alpha-1)(1-y)]
\label{fd rTL}
\end{equation}
with $0<\alpha\leq 2, \nu>0$ and $0\leq y\leq 1$. The density of the \textit{rGTL} can be strictly decreasing, strictly increasing or may possess a mode or an anti-mode, according to the values of the parameters.

In order to define the new distribution, we consider a sequence of independent and identically distributed continuous random variables $Y_1, Y_2, ...Y_Z$, with distribution function $G(y)$, where $Z$ is a discrete random variable following a Power Series distribution truncated at zero, with probability function (\textit{pf}) given by
\begin{equation}
P(Z=z)=\frac{a_z \theta^z}{A(\theta)}, \ \ z=1,2,...; \ \ \theta>0
\end{equation}
where $A(\theta)=\sum_{z=1}^\infty a_z\theta^z$ is finite and $a_z \geq 0$. In Table \ref{PS Distr} the most common distributions belonging to the Power Series family are reported.

\begin{table}
\centering
\begin{tabular}{rcccc}
\hline
Distribution &     $a_z$ & $A(\theta)$ & $A'(\theta)$ & range of $\theta$\\ 
\hline
 Logarithmic &     $1/z $ & $-\log(1-\theta)$ & $(1-\theta)^{-1}$ &   $(0,1)$ \\

 Geometric &       $1$ & $\theta(1-\theta)^{-1}$ & $(1-\theta)^{-2}$ & $(0,1)$ \\

   Poisson &    $1/z!$ & $e^{\theta}-1$ & $e^{\theta}$ &  $(0,\infty)$  \\

  Binomial & $\binom {m} {z}$ & $(\theta+1)^m-1$ & $m(\theta+1)^{m-1}$ & $(0,\infty)$ \\
\hline
\end{tabular}
\caption{Some special cases of the Power Series distribution.}
\label{PS Distr}
\end{table}

Let $Y=\min \{Y_1,...Y_Z\}$. The conditional \textit{pdf} of $Y$ given that $Z=z$, i.e. $Y|Z=z$,  can be obtained from the distribution of the minimum of $z$ random variables:
\begin{equation}
f(y|Z=z)=z \cdot  [1-G(y)]^{z-1} \cdot  g(y).
\end{equation}
Hence, the joint \textit{pdf} of $(Y,Z)$ is given by 
\begin{equation}
f(y, z)=f(y|Z=z)P(Z=z)=\frac{a_z \theta^z}{A(\theta)} z \cdot [1-G(y)]^{z-1} \cdot  g(y).
\label{jfd}
\end{equation}

By denoting with $A'(\cdot)$ the derivative of $A(\cdot)$ with respect to the argument, the marginal \textit{pdf} of $Y$ is
\begin{eqnarray}
f(y)=\sum_{z=1}^{\infty} \frac{a_z \theta^z}{A(\theta)} z \cdot  [1-G(y)]^{z-1} \cdot  g(y)=\frac{\theta g(y)A'\{\theta  [1-G(y)]\}}{A(\theta)},
\label{fd_comp}
\end{eqnarray}
with \textit{cdf} given by
\begin{equation}
F(y)=1-\sum_{z=1}^{\infty} \frac{a_z \theta^z}{A(\theta)}  \cdot [1-G(y)]^{z} =1-\frac{A\{\theta [1-G(y)]\}}{A(\theta)}.
\label{fr_comp}
\end{equation}

Replacing the expressions (\ref{fdr rTL}) and (\ref{fd rTL}) in (\ref{fd_comp}) and (\ref{fr_comp}), we obtain the \textit{rGTL-PS} \textit{pdf} and the corresponding cumulative \textit{cdf} as follows:
\begin{eqnarray}
f(y;\alpha,\nu,\theta)&=& \sum_{z=1}^{\infty}\frac{a_z \theta^z}{A(\theta)} z \nu\left[\alpha- 2(\alpha-1)(1-y)\right]\left( 1-y\right)^{\nu z-1} \left[\alpha-(\alpha-1)(1-y)\right]^{\nu z-1}\nonumber \\
&=&\sum_{z=1}^{\infty}\frac{a_z \theta^z}{A(\theta)} g(y;\alpha,\nu z),
\label{fd_rGTLPS}
\end{eqnarray}
\begin{eqnarray}
F(y;\alpha,\nu,\theta)&=&1-\sum_{z=1}^{\infty}\frac{a_z \theta^z}{A(\theta)} (1-y)^{\nu z} \left[\alpha- (1-\alpha) (1-y)\right]^{\nu z} \nonumber \\
&=&1-\sum_{z=1}^{\infty}\frac{a_z \theta^z}{A(\theta)} [1-G(y;\alpha, \nu z)]=\sum_{z=1}^{\infty}\frac{a_z \theta^z}{A(\theta)} G(y;\alpha, \nu z).
\label {fdr_rGTLPS}
\end{eqnarray}

It is evident, from (\ref{fd_rGTLPS}), that the density of the \textit{rGTL-PS } class can be expressed as mixture of \textit{rGTL} densities, $g(y;\alpha, \nu z)$, with weights $\frac{a_z \theta^z}{A(\theta)}$. In the following, we refer to this property because it enables us to obtain some mathematical properties of the \textit{rGTL-PS} distributions, such as the moments.

\section{Statistical Properties}
In this section, we study some properties of the \textit{rGTL-PS} distribution. In particular, we determine 
the \textit{rth} incomplete moment and ordinary moments using the known properties of the mixture distributions. Moreover, we compute the quantile and the hazard rate, evaluating its behaviour at the extrems of the support.

\subsection{Moments}

In order to calculate the \textit{rth} moment of \textit{rGTL-PS} distribution, first we determine the incomplete moment of order \textit{r} for a \textit{rGTL} distribution then, by using the properties of the mixture distributions, we calculate the \textit{rth} moment of a \textit{rGTL-PS} distribution.

\begin{Lemma}
If $Y\sim rGTL(\alpha, \nu)$ then the incomplete moment of order $r$ is \\
\begin{eqnarray*}
E\left\{Y^r|Y\leq y^{*}\right\}&=&\alpha^{\nu-1}\Gamma(\nu+1)\sum^{\infty}_{h=0} \frac{(-k)^h}{\Gamma(\nu -h)h!} \\
&& \left\{(2-\alpha)B_{y^{*}}\left( r+1, \nu+h \right)+2(\alpha-1)B_{y^{*}}\left( r+2, \nu+h \right) \right\}
\end{eqnarray*}
where $y^{*} \in (0,1]$ and $B_{y^{*}}(a, b)=\int^{y^{*}}_{0} y^{a-1}(1-y)^{b-1}dy$.
\end{Lemma}
textbf{\textit{Proof}}
Using $(1-kz)^{\nu-1}=\sum^{\infty}_{h=0} \frac{(-1)^{h} \Gamma(\nu) (kz)^{h} }{\Gamma(\nu-h)h!}$, with $k=\frac{\alpha-1}{\alpha}$ we can write
\begin{eqnarray*}
E\left\{Y^r|Y\leq y^{*}\right\}&=& \nu \alpha^{\nu-1}  \left\{ (2-\alpha) \int^{y^{*}}_{0} w^{r} (1-w)^{\nu-1} \left[1-k(1-w)\right]^{\nu-1} dw \right. \\
  && \left. +2(\alpha-1) \int^{y^{*}}_{0} w^{r+1} (1-w)^{\nu-1} \left[ 1-k(1-w)\right]^{\nu-1} dw  \right\} \\
	&=& \nu \alpha^{\nu-1} \sum^{\infty}_{h=0} \frac{\Gamma(\nu)(-k)^h}{\Gamma(\nu -h)h!} \left\{ (2-\alpha) \int^{y^{*}}_{0} w^{r} (1-w)^{\nu+h-1} dw \right\} \\
	&& \left. + 2(\alpha-1) \int^{y^{*}}_{0} w^{r+1} (1-w)^{\nu+h-1} dw \right\}. 
\end{eqnarray*} 
\begin{Cor} If $Y\sim rGTL(\alpha, \nu)$ then the moment of order $r$ is
\begin{eqnarray}
E\left\{Y^r\right\}= \alpha^{\nu-1} \Gamma(\nu+1) \sum^{\infty}_{h=0} \frac{(-k)^h}{\Gamma(\nu -h)h!} \left[(2-\alpha)+
\frac{2(\alpha-1)(r+1)} {(\nu+h+r+1)} \right]B(r+1, \nu+h).
\label{mom_rGTL}
\end{eqnarray}
\end{Cor}

\textbf{\textit{Proof}}
 It is enough to put $y^{*}=1$ in the \textit{Lemma}. 


From (\ref{mom_rGTL}), it is easy to obtain the moment of order $r$ of the \textit{rGTL-PS} distribution:
\begin{eqnarray*}
E(Y^r) &=&\sum_{z=1}^{\infty} \frac{a_z \theta^z}{A(\theta)} E_{rGTL}(Y^r;\alpha,\nu z)=\nonumber  \\
&=&\sum_{z=1}^{\infty} \frac{a_z \theta^z}{A(\theta)} \nu z \alpha^{\nu z-1} \sum^{\infty}_{h=0} \frac{(-k)^h}{\Gamma(\nu z -h)h!} \left[(2-\alpha)+
\frac{2(\alpha-1)(r+1)} {(\nu z+h+r+1)} \right]B(r+1, \nu z+h). \\
\label{mom_rGTLPS}
\end{eqnarray*}

\subsection{Hazard rate}
In this section, we verify that the hazard rate of \textit{rGTL-PS} distribution is more flexible than the hazard rate of the \textit{rGTL}.

First of all, it is easy to show that the hazard rate of \textit{rGTL} distribution is always increasing. Indeed, by (\ref{fdr rTL}) and (\ref{fd rTL}) we obtain the hazard rate for the \textit{rGTL}, as follows:
\begin{eqnarray}
h_{rGTL}(y;\alpha,\nu)=\nu \frac{\alpha-2(\alpha-1)(1-y)}{\alpha(1-y)-(\alpha-1)(1-y)^2 }
\label{hz_rGTL}
\end{eqnarray}
and, after easy algebra, the derivative of (\ref{hz_rGTL}) with respect to $y$ is given by
\begin{equation*}
\frac{\partial h_{rGTL}(y;\alpha,\nu)}{\partial y}=\frac{\nu}{[\alpha(1-y)-(\alpha-1)(1-y)^2]^2}\cdot \left\{[(\alpha-1)(1-y)-\alpha]^2+(\alpha-1)^2(1-y)^2  \right\}
\end{equation*}
that is positive $\forall y $, $\nu>0$ and $\alpha \in (0,2] $.

Starting from (\ref{fd_comp}) and (\ref{fr_comp}), we obtain the expression for the hazard rate of the \textit{rGTL-PS} distribution, as follows:
\begin{eqnarray}
h(y;\alpha,\nu,\theta)=\frac{\theta \cdot g(y;\alpha,\nu) A'\{\theta [1-G(y;\alpha,\nu)]\}}{A\{\theta [1-G(y;\alpha,\nu)]\}}. 
\label{hz_rGTLPS}
\end{eqnarray}

The hazard rate is a complex function of $y$. However, from (\ref{hz_rGTLPS}) we have that for  $y \rightarrow 0$, the hazard rate tends to $\nu\theta (2-\alpha)\cdot A'(\theta)/A(\theta)$, while for $y\rightarrow 1$ the hazard rate tends to $+\infty$. Indeed, we have:
\begin{eqnarray}
\lim_{y \rightarrow 1} h(y;\alpha,\nu,\theta)&=&\lim_{y \rightarrow 1}\frac{ A'\{ \theta [1-G(y;\alpha,\nu)]\}}{A\{\theta [1-G(y;\alpha,\nu)]\}} \theta g(y;\alpha,\nu)\nonumber\\
&=&a_1 \theta  \lim_{y \rightarrow1} \frac{g(y;\alpha,\nu)}{A\{\theta [1-G(y;\alpha,\nu)]\}}
\end{eqnarray}
where $A'(\theta)=\frac{\partial A(\theta)}{\partial \theta}$ has the same radius of convergence of $A(\theta)$ and $A'(0)=a_1>0$.  Observed that $A(0)=0$, if $\nu\leq 1$ then $\lim_{y \rightarrow 1} h(y;\alpha,\nu,\theta)=+\infty$, given that $\lim_{y \rightarrow 1} g(y;\alpha,\nu)=+\infty$; while, if $\nu>1$, given that $\lim_{y \rightarrow 1} g(y;\alpha,\nu)=0$, we have $\lim_{y \rightarrow 1} h(y;\alpha,\nu,\theta)=\frac{0}{0}$. Now, using l'Hopital's Rule we obtain:
\begin{eqnarray}
\lim_{y \rightarrow 1 } h(y;\alpha,\nu,\theta) &=&  a_1  \theta\lim_{y \rightarrow 1}\frac{ g'(y;\alpha,\nu)}{A'\{\theta[1-G(y;\alpha,\nu)]\}[-\theta g(y;\alpha,\nu)]}\nonumber \\
&=&- \lim_{y \rightarrow 1 } \frac{ g'(y;\alpha,\nu)}{ g(y;\alpha,\nu)}.
\end{eqnarray}
It can be shown that $\lim_{y \rightarrow 1}\frac{ g'(y;\alpha,\nu)}{g(y;\alpha,\nu)}=-\infty$, so that $\lim_{y \rightarrow 1}h(y;\alpha,\nu,\theta)=+\infty$.

As we can see from Fig. \ref{fig:casi}, the hazard rate for the models belonging to the \textit{rGTL-PS} class of distributions is much more flexible than the hazard rate of the \textit{rGTL} distribution. In particular, besides having the increasing shape, it is also possible to have the bathtub shape and the upside-down bathtub and then bathtub shape. This wide range of different behaviours of the hazard rate makes the models belonging to the \textit{rGTL-PS} class suitable models for reliability theory and survival analysis in cases of bounded domain.

\subsection{Quantile}
An advantage of the \textit{rGTL-PS} distribution is the possibility to get the expression for the quantile function. Indeed, the \textit{cdf} for the \textit{rGTL-PS} can be expressed in a closed form, as it can be noted from the expression (\ref{fr_comp}) and therefore the quantile can be easily obtained by remembering the expression for the quantile of \textit{rGTL} distribution:
\\
\begin{equation}
y_p=\begin{cases}
1-\frac{\alpha-\sqrt{\alpha^2-4(\alpha-1)(1-p)^{1/\nu}}}{2(\alpha-1)} & 1<\alpha\leq 2 \\
1-(1-p)^{1/\nu} & \alpha=1\\
1-\frac{\alpha+\sqrt{\alpha^2-4(\alpha-1)(1-p)^{1/\nu}}}{2(\alpha-1)} & 0\leq\alpha\leq 1
\end{cases}
\end{equation}
\\
and by putting $q=F(y_q;\alpha,\nu,\theta)=1-\frac{A\{\theta [1-G(y_q;\alpha,\nu)]\}}{A(\theta)}$.

Thus, the quantile $y_q$ of the \textit{rGTL-PS} distribution is given by
\begin{eqnarray}
y_q&=&G^{-1}\left(1-\frac{A^{-1}[(1-q)A(\theta)]}{\theta};\alpha,\nu\right) =
\nonumber\\
&=&\begin{cases}
1-\frac{\alpha-\sqrt{\alpha^2-4(\alpha-1)\left(\frac{A^{-1}[(1-q)A(\theta)]}{\theta}\right)^{1/\nu}}}{2(\alpha-1)} & 1<\alpha\leq 2 \\
1-\left(\frac{A^{-1}[(1-q)A(\theta)]}{\theta}\right)^{1/\nu} & \alpha=1\\
1-\frac{\alpha+\sqrt{\alpha^2-4(\alpha-1)\left(\frac{A^{-1}[(1-q)A(\theta)]}{\theta}\right)^{1/\nu}}}{2(\alpha-1)} & 0\leq\alpha\leq 1.
\end{cases}
\label{quantile}
\end {eqnarray}


\section{Inference}
In order to estimate the parameters $\bm{\eta} = (\alpha, \nu,\theta)$ of the \textit{rGTL-PS} distribution, we consider the maximum likelihood (\textit{ML}) method. Let $y = (y_1, y_2, ... ,y_n)$ be a random sample of size $n$ from the  \textit{rGTL-PS} given by (\ref{fd_comp}). The log-likelihood function for the vector of parameters $\eta = (\alpha, \nu,\theta)$  can be expressed as:
\begin{eqnarray}
\ell(\bm{\eta};\bm{y})=n \ln \theta - n \ln A(\theta) + \sum_{i=1}^n \ln g(y_i; \alpha, \nu) + \sum_{i=1}^n \ln A^{'}\left\{ \theta \left[1-G(y_i; \alpha, \nu) \right]\right\}.
\end{eqnarray}
Differentiating $\ell(\bm{\eta};\bm{y})$ with respect to $\alpha, \nu$ and $\theta$, respectively, and setting the results equal to zero, we have:
\begin{eqnarray}
\begin{cases}
\frac{\partial \ell(\bm{\eta};\bm{y})}{\partial \alpha}=\sum_{i=1}^n \frac{\dot{g}_{\alpha}(y_i; \alpha, \nu) }{g(y_i; \alpha, \nu)} -
\theta \sum_{i=1}^n \frac{ A^{''}\left\{ \theta \left[1-G(y_i; \alpha, \nu) \right]\right\} }{  A^{'}\left\{ \theta \left[1-G(y_i; \alpha, \nu) \right]\right\} }
\dot{G}_{\alpha}(y_i; \alpha, \nu)=0 \\
\frac{\partial \ell(\bm{\eta};\bm{y})}{\partial \nu}=\sum_{i=1}^n \frac{\dot{g}_{\nu}(y_i; \alpha, \nu) }{g(y_i; \alpha, \nu)} -
\theta \sum_{i=1}^n \frac{ A^{''}\left\{ \theta \left[1-G(y_i; \alpha, \nu) \right]\right\} }{  A^{'}\left\{ \theta \left[1-G(y_i; \alpha, \nu) \right]\right\} }
\dot{G}_{\nu}(y_i; \alpha, \nu)=0 \\
\frac{\partial \ell(\bm{\eta};\bm{y})}{\partial \theta}=\frac{n}{\theta}-\frac{A^{'}(\theta)}{A(\theta)}+
\sum_{i=1}^n \frac{A^{''}\left\{ \theta \left[1-G(y_i; \alpha, \nu) \right]\right\}}{A^{'}\left\{ \theta \left[1-G(y_i; \alpha, \nu) \right]\right\}}
\left[1-G(y_i; \alpha, \nu) \right]=0
\end{cases}
\end {eqnarray}
where the quantities $\dot{g}(.)$ and $\dot{G}(.)$ are reported in the \textit{Appendix}.
The system does not admit any explicit solution; therefore, the \textit{ML} estimates $\hat{\bm{\eta}} = (\hat{\alpha}, \hat{\nu},\hat{\theta})$  can only be obtained by means of numerical procedures. Under the usual regularity conditions, the
known asymptotic properties of the maximum likelihood method ensure that $\sqrt n(\hat{\bm{\eta}}_n-\bm{\eta})\stackrel{d} \rightarrow N(\bm{0},\hat{\bm{\Sigma}}_{\bm{\eta}})$, where $\hat{\bm{\Sigma}}_{\bm{\eta}}=[\lim_{n\rightarrow \infty}\bm{I}(\bm{\eta})/n]^{-1}$ is the asymptotic variance-covariance matrix and $\bm{I}(\bm{\eta})$ is the Fisher Information matrix.
Moreover, the asymptotic variance-covariance matrix of $\hat{\bm{\eta}}$ can be approximated by the inverse of the observed information matrix $\bm {J}(\bm{\eta})$, whose entries are given in the \textit{Appendix}.
In order to build the confidence intervals and hypothesis tests, we use the fact that the asymptotic distribution of $\boldsymbol{\hat{\eta}}$ can be approximated by the multivariate normal distribution, $N_{3}(\boldsymbol{\eta},\left[\textbf{J}(\boldsymbol{\hat{\eta}})\right]^{-1} )$, where $\left[\textbf{J}(\boldsymbol{\hat{\eta}})\right]^{-1}$ is the inverse of the observed information matrix evaluated in $\boldsymbol{\hat{\eta}}$.

\subsection{EM algorithm}

In the original formulation due to  Dempester et al. (1977), \textit{EM} algorithm is a method for computing maximum likelihood estimate iteratively, starting from some initial guess, when the data are incomplete. The \textit{EM} algorithm has become a popular tool in the statistical estimation problems involving incomplete data, or in problems which can be posed in similar form.
Each iteration of the \textit{EM} algorithm consists of an Expectation (\textit{E}) step, in which we calculate conditional expectation of the complete-data log-likelihood function given observed data, and a Maximization (\textit{M}) step, in which we maximize this equation. \\
In our case, we suppose that the complete-data $W_i=(Y_i,Z_i)$ ($ i=1,...n$) consist of an observable part $Y_i$ and an unobservable part $Z_i$. By (\ref{jfd}) the corresponding log-likelihood function \\
\begin{equation}
 \ell^{*}(\bm{\eta};\textbf{w}) \propto \ln(\theta)\sum^{n}_{i=1} z_{i}- \ln A(\theta) + \sum^{n}_{i=1} (z_{i}-1) \ln[1-G(y_i;\alpha, \nu)] + \sum^{n}_{i=1} \ln g(y_i;\alpha, \nu).
\label{ cdLike }
 \end{equation}
Then, the \textit{E}-step of the algorithm requires the computation of the conditional expectation
\begin{equation}
E[\ell^{*}(\bm{\eta}^{(r+1)};\textbf{W})| \textbf{Y}=\textbf{y}, \bm{\eta}^{(r)}]
\label{ condExp }
 \end{equation}
where $\bm{\eta}^{(r)}=(\alpha^{(r)}, \nu^{(r)}, \theta^{(r)})$ is the current estimate of $\bm{\eta}$ in the \textit{rth} iteration. It is straightforward to verify that the \textit{E}-step of an \textit{EM} cycle requires the computation of the conditional expectation of the conditional random variable $(Z|Y=y, \bm{\eta}^{(r)})$. Using  (\ref{jfd}) and (\ref{fd_comp}), the conditional probability function of $(Z|Y=y, \bm{\eta}^{(r)})$ is
\begin{equation}
g(z|y; \bm{\eta}) = \frac{a_{z} z \theta^{z-1} [1-G(y;\alpha, \nu)]^{z-1} }{ A ^{'}\left\{ \theta [1-G(y;\alpha, \nu)]\right\} }
\label{ condpf }
\end{equation}
for $z\in \aleph$ and $y\in (0,1)$. After easy algebra, the conditional expectation is given by
\begin{equation}
E[Z|Y=y, \bm{\eta}^{(r)}  ] = \theta^{(r)}
\frac{  A ^{''} \left\{ \theta^{(r)} [1-G(y;\alpha^{(r)}, \nu^{(r)})] \right\} }{ A ^{'} \left\{\theta^{(r)} [1-G(y;\alpha^{(r)}, \nu^{(r)})]\right\}  } + 1.
\label{ condExprTL }
\end{equation}
The \textit{M}-step of \textit{EM} algorithm requires the maximization of the complete-data likelihood over $\bm{\eta}$, with the unobservable part $z_i$ replaced by their conditional expectations. Thus the $\bm{\eta}^{(r+1)} $ at $\textit{(r+1)th}$ of \textit{EM} is the numerical solution of the following nonlinear system:
\begin{eqnarray}
\begin{cases}
\frac{\partial \ell^{*}(\bm{\eta}^{(r)};\textbf{W})}{\partial \alpha}=-\sum_{i=1}^n (z_i -1)
\frac{\dot{G}_{\alpha} \left( y_i;\alpha^{(r)}, \nu^{(r)} \right) }{ \left[1-G_{\alpha}\left( y_i;\alpha^{(r)}, \nu^{(r)} \right) \right]}
+\sum_{i=1}^n \frac{\dot{g}_{\alpha}\left( y_i;\alpha^{(r)}, \nu^{(r)} \right) }{ \left[1-g_{\alpha}\left( y_i;\alpha^{(r)}, \nu^{(r)} \right)\right]} =0 \\
\frac{\partial \ell^{*}(\bm{\eta}^{(r)};\textbf{W})}{\partial \nu}=-\sum_{i=1}^n (z_i -1)
\frac{\dot{G}_{\nu} \left( y_i;\alpha^{(r)}, \nu^{(r)} \right) }{ \left[1-G_{\nu}\left( y_i;\alpha^{(r)}, \nu^{(r)} \right) \right]}
+\sum_{i=1}^n \frac{\dot{g}_{\nu}\left( y_i;\alpha^{(r)}, \nu^{(r)} \right) }{ \left[1-g_{\nu}\left( y_i;\alpha^{(r)}, \nu^{(r)} \right)\right]} =0 \\
\frac{\partial \ell^{*}(\bm{\eta}^{(r)};\textbf{W})}{\partial \theta}=\frac{1}{\theta^{(r)}}\sum_{i=1}^n z_{i} -
n \frac{A^{'}(\theta ^{(r)}) }{A(\theta ^{(r)}) }=0.
\end{cases}
\end {eqnarray}

\section{Special cases}
\begin{figure}
 \centering
 \includegraphics[width=0.38\textwidth]{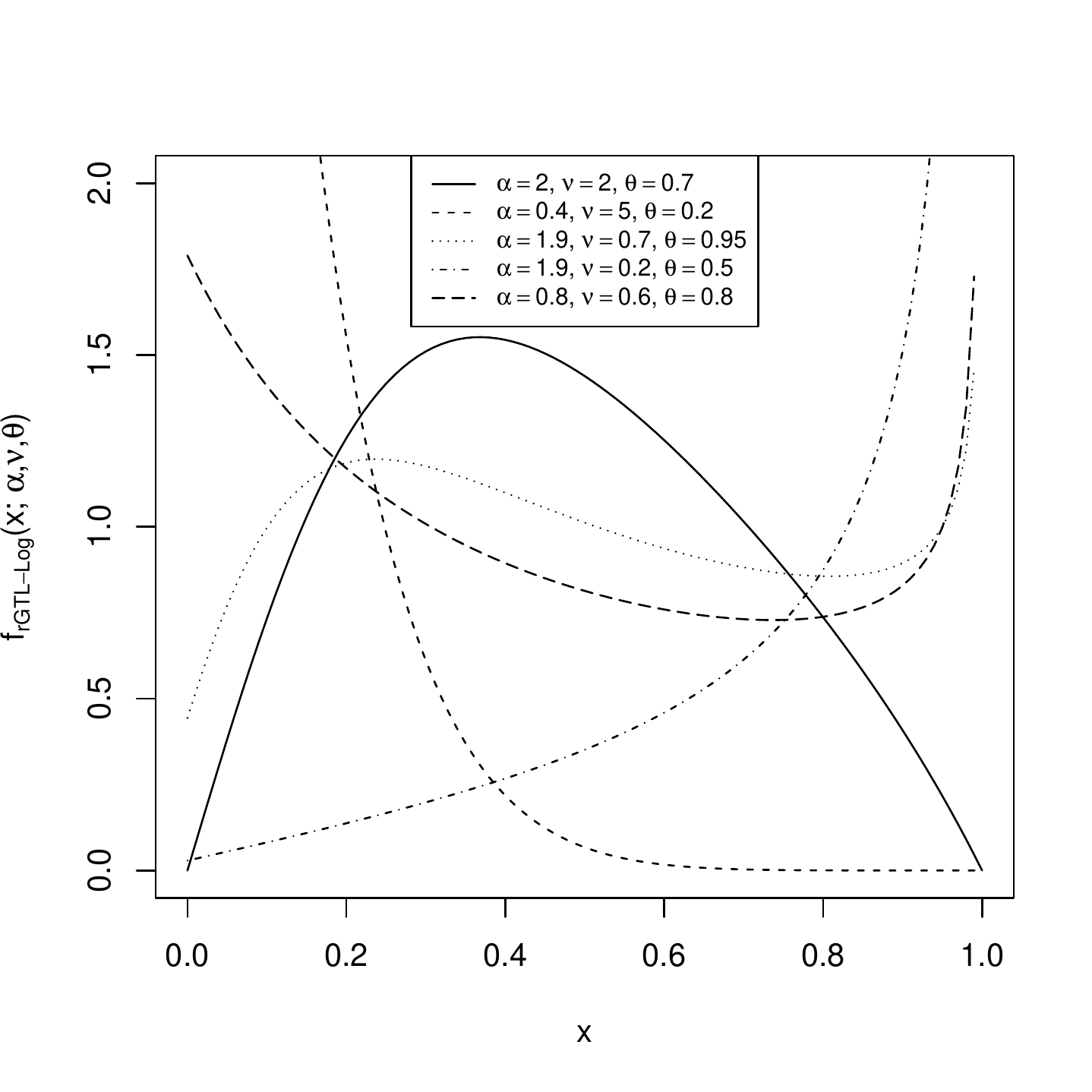} \includegraphics[width=0.38\textwidth]{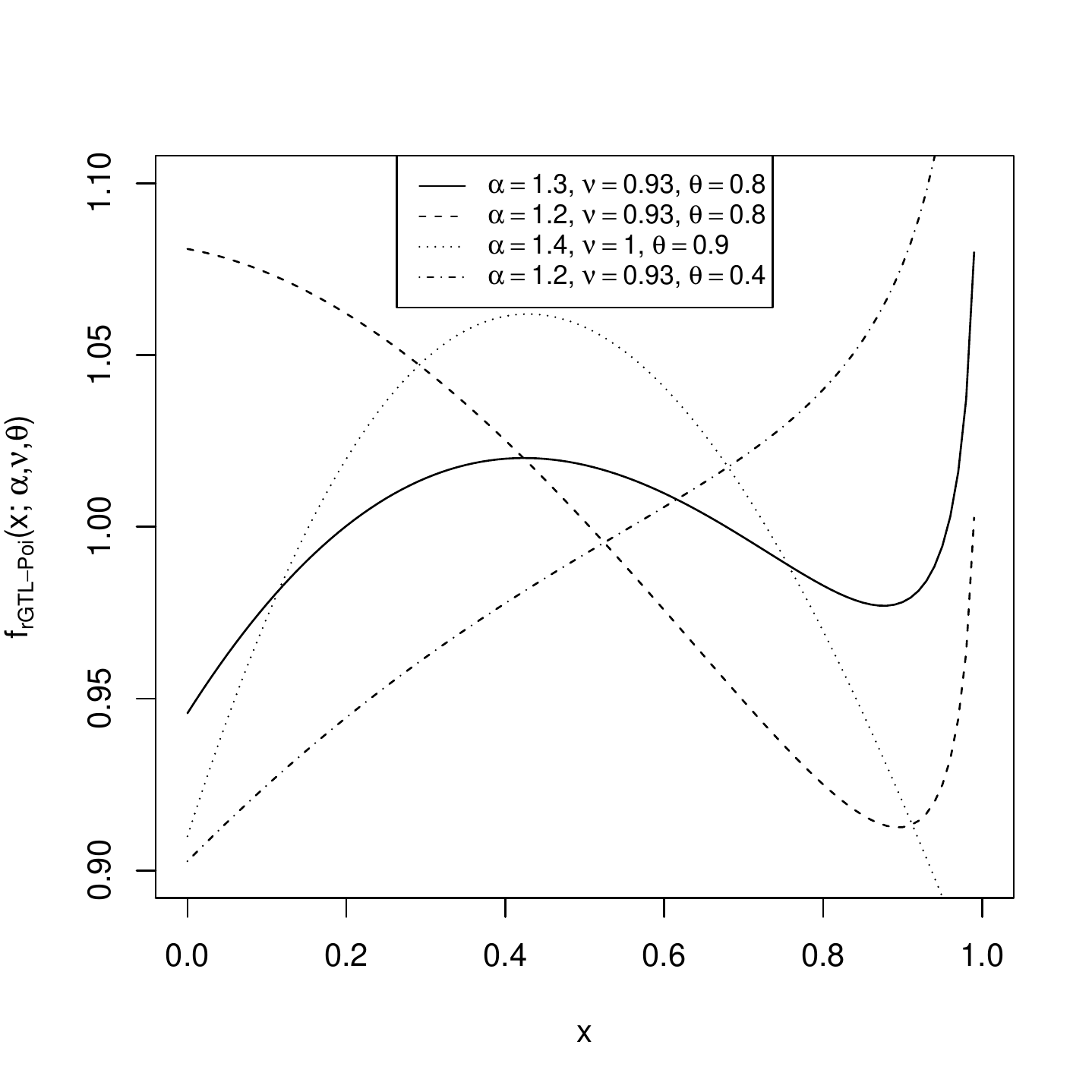}\\ 
 \includegraphics[width=0.38\textwidth]{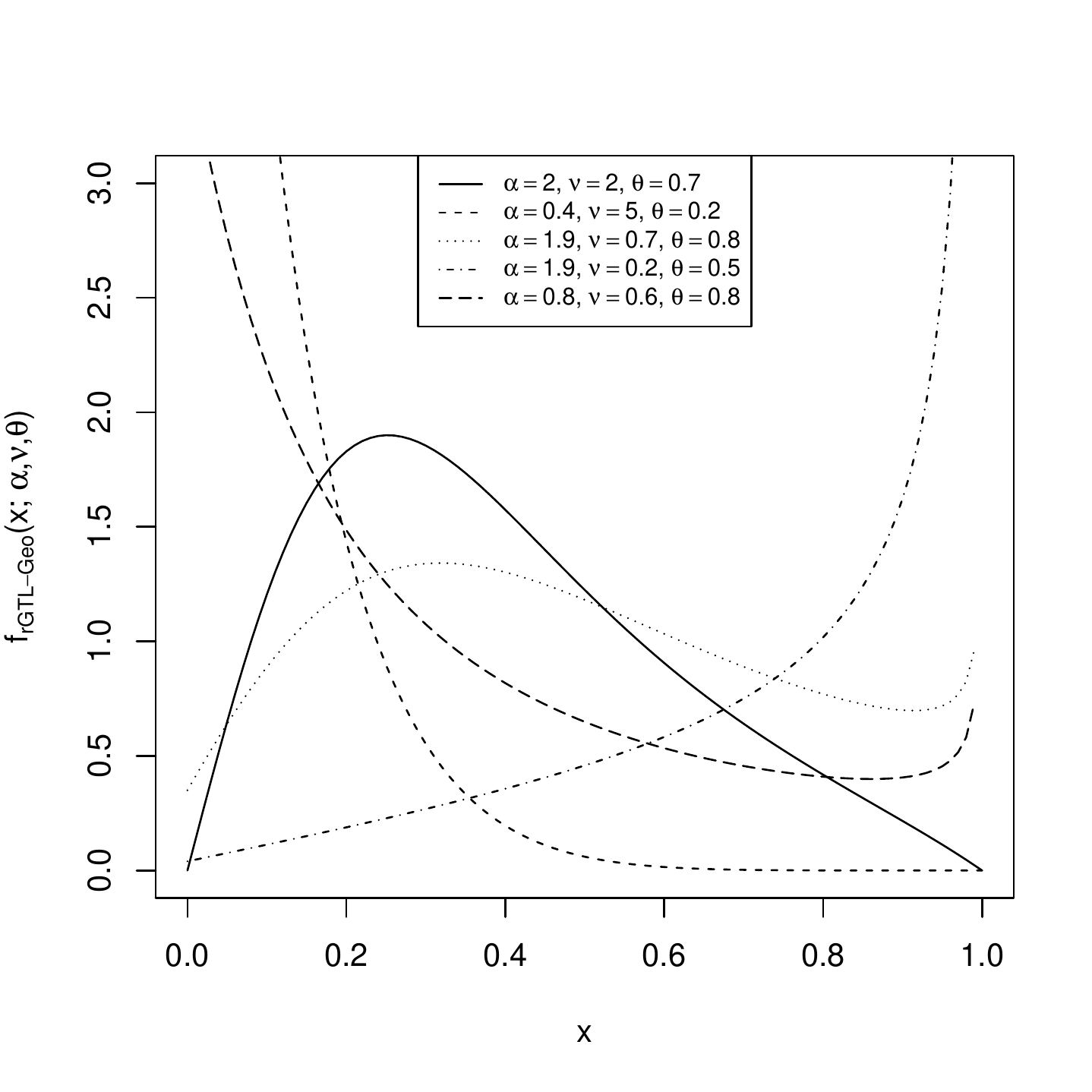} \includegraphics[width=0.38\textwidth]{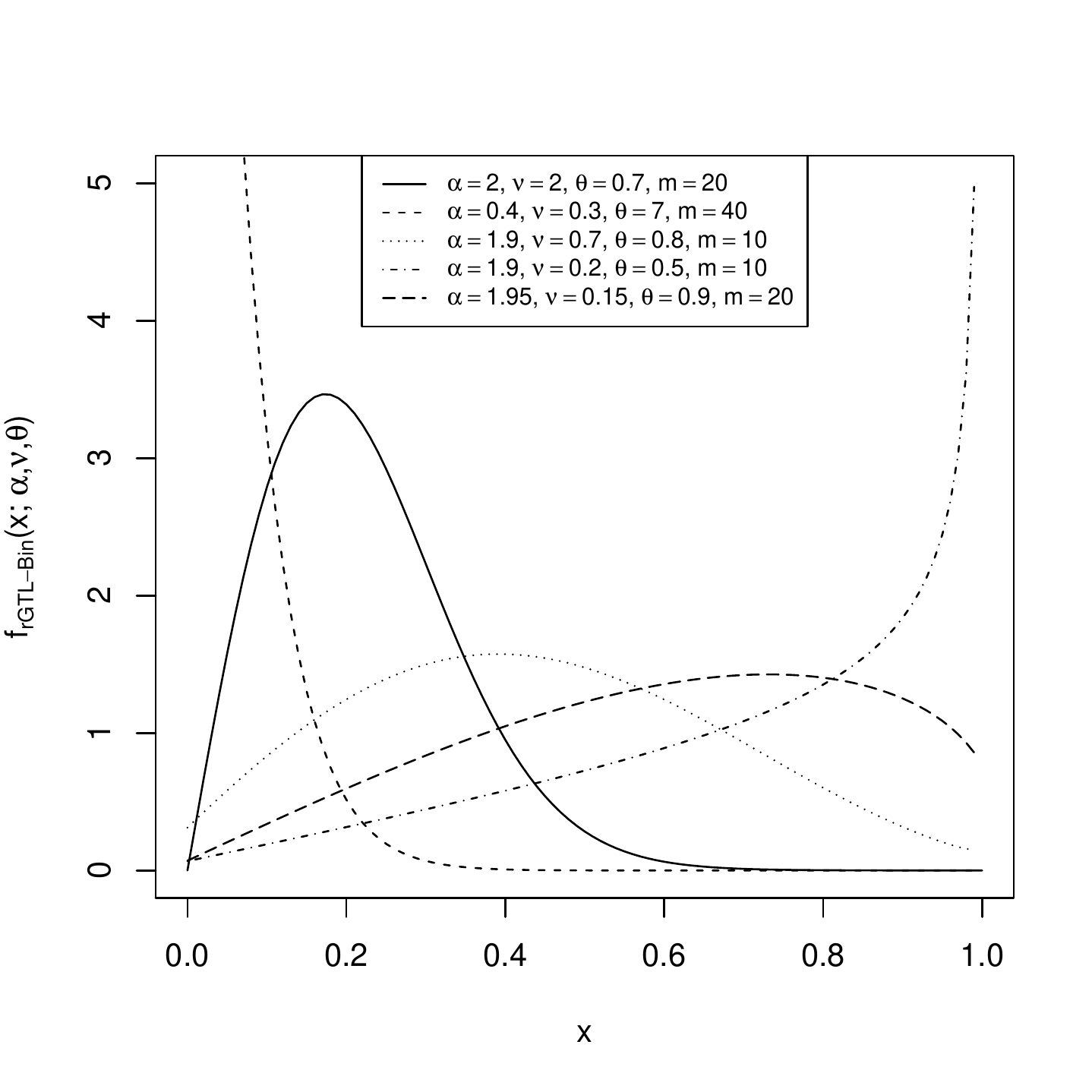}\\
 \includegraphics[width=0.38\textwidth]{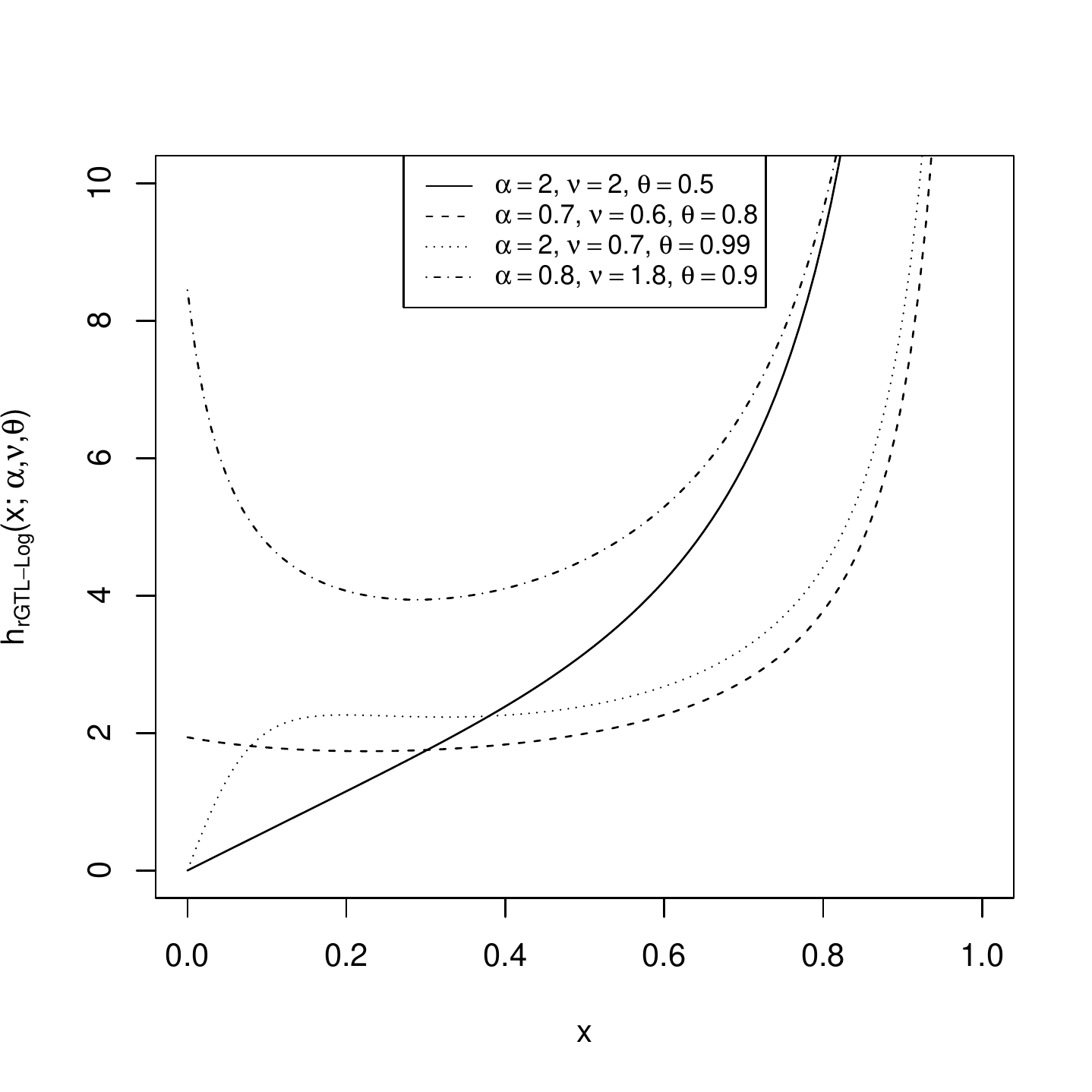} \includegraphics[width=0.38\textwidth]{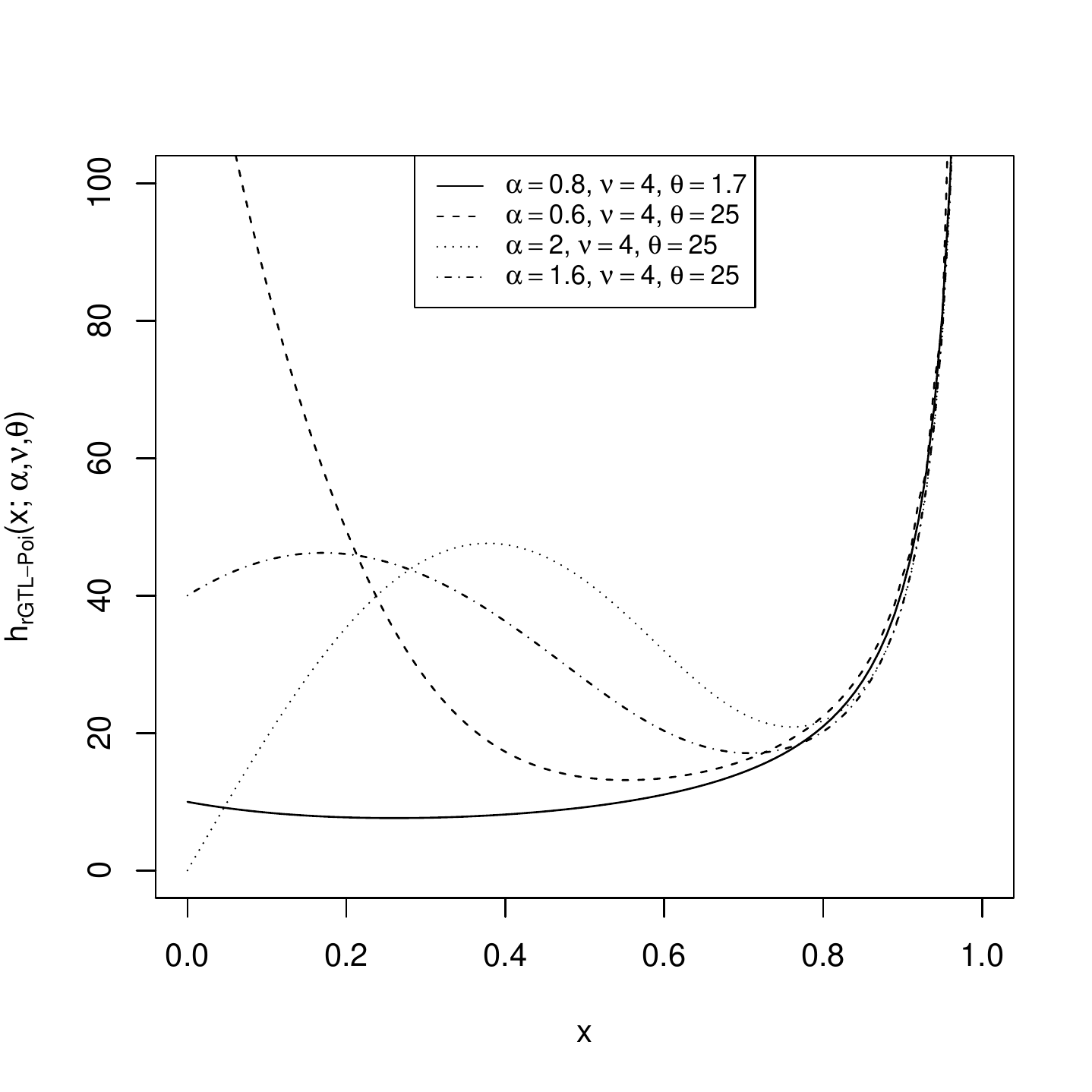}\\
\includegraphics[width=0.38\textwidth]{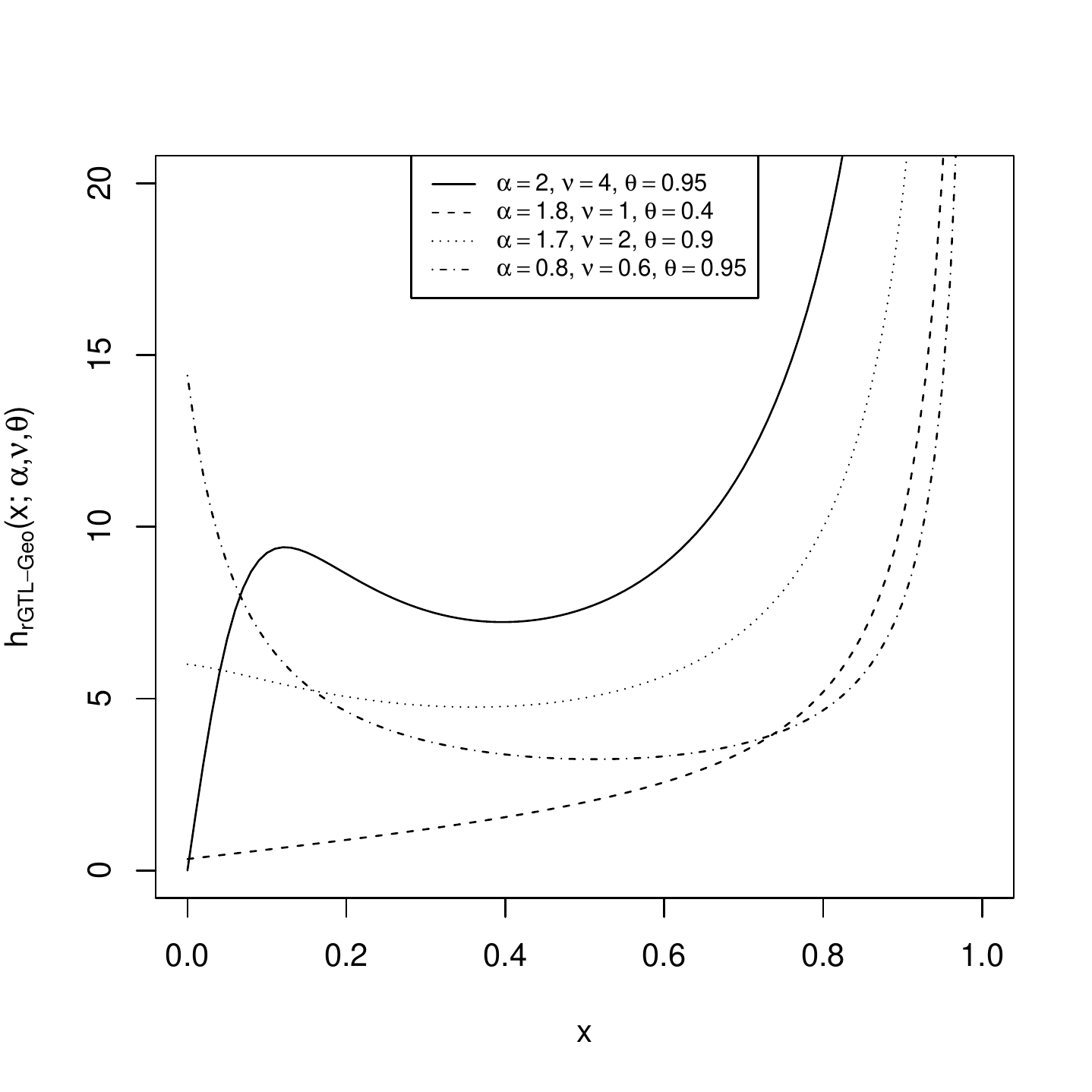} \includegraphics[width=0.38\textwidth]{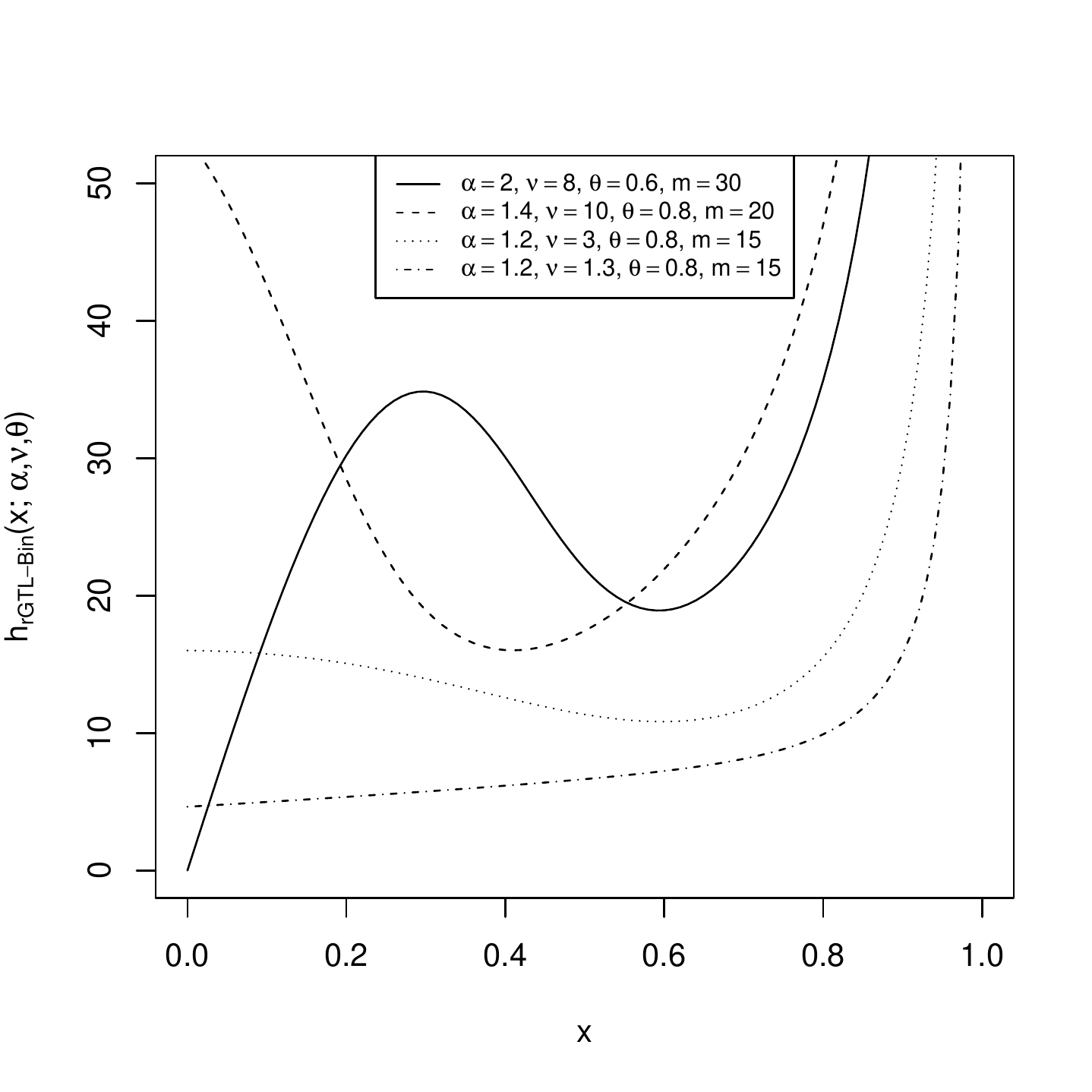}\\
  \caption{Density and hazard rate for some models of \textit{rGTL-PS} class for certain parameter values.}
\label{fig:casi}
\end{figure}

In this section, we furnish some results about special cases of the \textit{rGTL-PS} class of distributions. In particular, by considering the quantities reported in Table \ref{PS Distr} and the expressions given in (\ref{fd_comp}) and (\ref{fr_comp}), we obtain the \textit{pdf} and the \textit{cdf} for \textit{rGTL}-Logarithmic,  \textit{rGTL}-Geometric, the \textit{rGTL}-Poisson and the \textit{rGTL}-Binomial distributions. Some plots of density functions and hazard rates are given in Fig. \ref{fig:casi} to show the flexibility of these models.

\subsection{\textit{rGTL}-Logarithmic distribution}
The \textit{pdf} and the \textit{cdf} of the \textit{rGTL}-Logarithmic (\textit{rGTL-Log}) random variable, respectively, are:
\begin{equation}
f(y;\alpha,\nu,\theta)=\frac{\theta \nu (1-y)^{\nu-1}[\alpha-(\alpha-1)(1-y)]^{\nu-1} [\alpha-2(\alpha-1)(1-y)]}{-\log(1-\theta)\{1-\theta(1-y)^{\nu}[\alpha-(\alpha-1)(1-y)]^\nu\}}
\label{fd rGTLLog}
\end{equation}

\begin{equation}
F(y;\alpha,\nu,\theta)=1-\frac{\log\{1-\theta(1-y)^\nu[\alpha-(\alpha-1)(1-y)]^\nu\}}{\log(1-\theta)}
\label{fdr rGTLLog}
\end{equation}

with $0<\alpha\leq 2, \nu>0$ and $0<\theta<1$. The hazard rate is given by
\begin{equation}
h(y;\alpha,\nu,\theta)=-\frac{\theta g(y;\alpha,\nu)}{[1-\theta G(y;\alpha,\nu)]\cdot[\log(1-\theta G(y;\alpha,\nu))]}.
\label{hz rGTLLog}
\end{equation}
Some plots of the \textit{pdf} and hazard rate are given in Fig. \ref{fig:casi}.
Finally, considering that $A^{-1}[(1-q)A(\theta)]=1-(1-\theta)^{1-q}$, from (\ref{quantile}), we obtain the quantile for \textit{rGTL-Log} distribution:
\begin{eqnarray}
y_q&=&\begin{cases}
1-\frac{\alpha-\sqrt{\alpha^2-4(\alpha-1)\left(\frac{1-(1-\theta)^{1-q}}{\theta}\right)^{1/\nu}}}{2(\alpha-1)} & 1<\alpha\leq 2 \\
1-\left(\frac{1-(1-\theta)^{1-q}}{\theta}\right)^{1/\nu} & \alpha=1\\
1-\frac{\alpha+\sqrt{\alpha^2-4(\alpha-1)\left(\frac{1-(1-\theta)^{1-q}}{\theta}\right)^{1/\nu}}}{2(\alpha-1)} & 0\leq\alpha\leq 1
\end{cases}
\end {eqnarray}

\subsection{\textit{rGTL}-Geometric distribution}
The \textit{pdf} and the \textit{cdf} of the \textit{rGTL}-Geometric (\textit{rGTL-Geo}) random variable, respectively, are:
\begin{equation}
f(y;\alpha,\nu,\theta)=\frac{\nu(1-\theta)(1-y)^{\nu-1}[\alpha-(\alpha-1)(1-y)]^{\nu-1}[\alpha-2(\alpha-1)(1-y)]}{\{1-\theta(1-y)^\nu[\alpha-(\alpha-1)(1-x)]^\nu\}^2}
\label{fd rGTLGeo}
\end{equation}

\begin{equation}
F(y;\alpha,\nu,\theta)=\frac{1-(1-y)^\nu[\alpha-(\alpha-1)(1-y)]^\nu}{1-\theta\{(1-y)^\nu[\alpha-(\alpha-1)(1-y)]^\nu\}}
\label{fdr rGTLGeo}
\end{equation}

with $0<\alpha\leq 2, \nu>0$ and $0<\theta<1$. After some passages we obtain the hazard rate, given by
\begin{equation}
h(y;\alpha,\nu,\theta)=\frac{ g(y;\alpha,\nu)}{[1- G(y;\alpha,\nu)]\cdot \{1-\theta [1-G(y;\alpha,\nu)]\}}.
\label{hz rGTLGeo}
\end{equation}
The quantile for the \textit{rGTL-Geo} distribution can be obtained from (\ref{quantile}), by considering that  $A^{-1}[(1-q)A(\theta)]=\frac{(1-q)\theta}{1-q\theta}$.

\subsection{\textit{rGTL}-Poisson distribution}
The \textit{pdf} and the \textit{cdf} of the \textit{rGTL}-Poisson (\textit{rGTL-Poi}) random variable, respectively, are:
\begin{eqnarray}
&&f(y;\alpha,\nu,\theta)=\nonumber\\
&&\frac{\nu \theta}{\exp(\theta)-1} \left[\alpha-2(\alpha-1) (1-y)\right] (1-y)^{\nu-1} \left[\alpha-(\alpha-1)(1-y)\right]^{\nu-1} \cdot \nonumber \\
&& \exp\{\theta(1-y)^{\nu} \left[\alpha- (\alpha-1)(1-y)\right]^{\nu}\}
\label{fd rGTLPoi}
\end{eqnarray}

\begin{equation}
F(y;\alpha,\nu,\theta)=1-\frac{\exp\{\theta (1-y)^{\nu} [\alpha-(1-\alpha)(1-y)]^{\nu}\}-1}{\exp(\theta)-1}
\label{fdr rGTLPoi}
\end{equation}
with $0<\alpha\leq 2, \nu>0$ and $\theta>0$. After some passages we obtain the hazard rate, given by
\begin{equation}
h(y;\alpha,\nu,\theta)=\frac{\theta g(y;\alpha,\nu) e^{\theta [1- G(y;\alpha,\nu)]}}{e^{\theta[1-G(y;\alpha,\nu)]}-1}.
\label{hz rGTLPoi}
\end{equation}

From the plots reported in Fig. \ref{fig:casi}, we can state that the hazard rate of the \textit{rGTL}-Poisson distribution can be monotonically increasing, bathtub and UB-BT.

In this case $A^{-1}[(1-q)A(\theta)]=\log[q(1-e^{\theta})+e^{\theta}]$, thus the quantile for the \textit{rGTL-Poi} distribution can be obtained by inserting this expression in (\ref{quantile}).

\subsection{\textit{rGTL}-Binomial distribution}
The \textit{pdf} and the \textit{cdf} of the \textit{rGTL}-Binomial (\textit{rGTL-Bin}) random variable, respectively, are:
\begin{eqnarray}
f(y;\alpha,\nu,\theta, m)&=&\frac{m \theta \nu(1-y)^{\nu-1}[\alpha-(\alpha-1)(1-y)]^{\nu-1}[\alpha-2(\alpha-1)(1-y)]}{(\theta+1)^m-1} \cdot \nonumber \\
&\cdot& \{\theta(1-y)^\nu[\alpha-(\alpha-1)(1-y)]^\nu+1\}^{m-1}
\label{fd rGTLBin}
\end{eqnarray}

\begin{equation}
F(y;\alpha,\nu,\theta, m)=\frac{(\theta+1)^m-\{\theta(1-y)^\nu [\alpha-(\alpha-1)(1-y)]^\nu+1\}^m-1}{(\theta+1)^m}
\label{fdr TLBin}
\end{equation}
with $0<\alpha\leq 2, \nu>0$, $\theta>0$ and $m=1,2,...$. After some passages we obtain the hazard rate, as:
\begin{equation}
h(y;\alpha,\nu,\theta,m)=\frac{m \theta g(y;\alpha,\nu) \{\theta [1- G(y;\alpha,\nu)]+1\}^{m-1}}{\{\theta[1-G(y;\alpha,\nu)]+1\}^m-1}.
\label{hz rGTLBin}
\end{equation}
Finally, from (\ref{quantile}), it is possible to obtain the expression for the quantile of \textit{rGTL-Bin} distribution, by considering that $A^{-1}[(1-q)A(\theta)]=\{(1+q)(\theta+1)^m-q\}^{1/m}-1$.

\section{Applications}
In this section, we fit some models, belonging to the proposed class, to real data. In particular, in the first example, we consider the dataset reported in \cite{Genc13}, regarding two different algorithms, SC16 and P3, used to estimate unit capacity factors by the electric utility industry, while, in the second examples, we consider the percentage of muslim population and the percentage of atheists, used by \citet{Silva14}.
\\

\textit{Example 1}. In \cite{Genc13}, the author fits the \textit{TL} distribution to the capacity data. We compare the reported results with those obtained considering the \textit{rGTL} distribution, the Beta distribution and three models from the \textit{rGTL-PS} class.

In Table \ref{stime} the \textit{ML} estiamates of the parameters, with the corresponding standard errors, and the values for Akaike Information Criterion (AIC)  are reported for both SC16 and P3 algorithms. The lower values of the AIC obtained for \textit{rGTL-Log} model, compared to those obtained in correspondence with all others models, suggest the superiority of the former in describing these data. Furthermore, Table \ref{stime} gives the results obtained from the Kolmogorov-Smirnov  (KS) test. Once again, the KS statistic for the \textit{rGTL-Log} distribution is the lowest among all those obtained for the considered distributions.

\begin{figure}[!ht]
 \centering
 \includegraphics[width=0.48\textwidth]{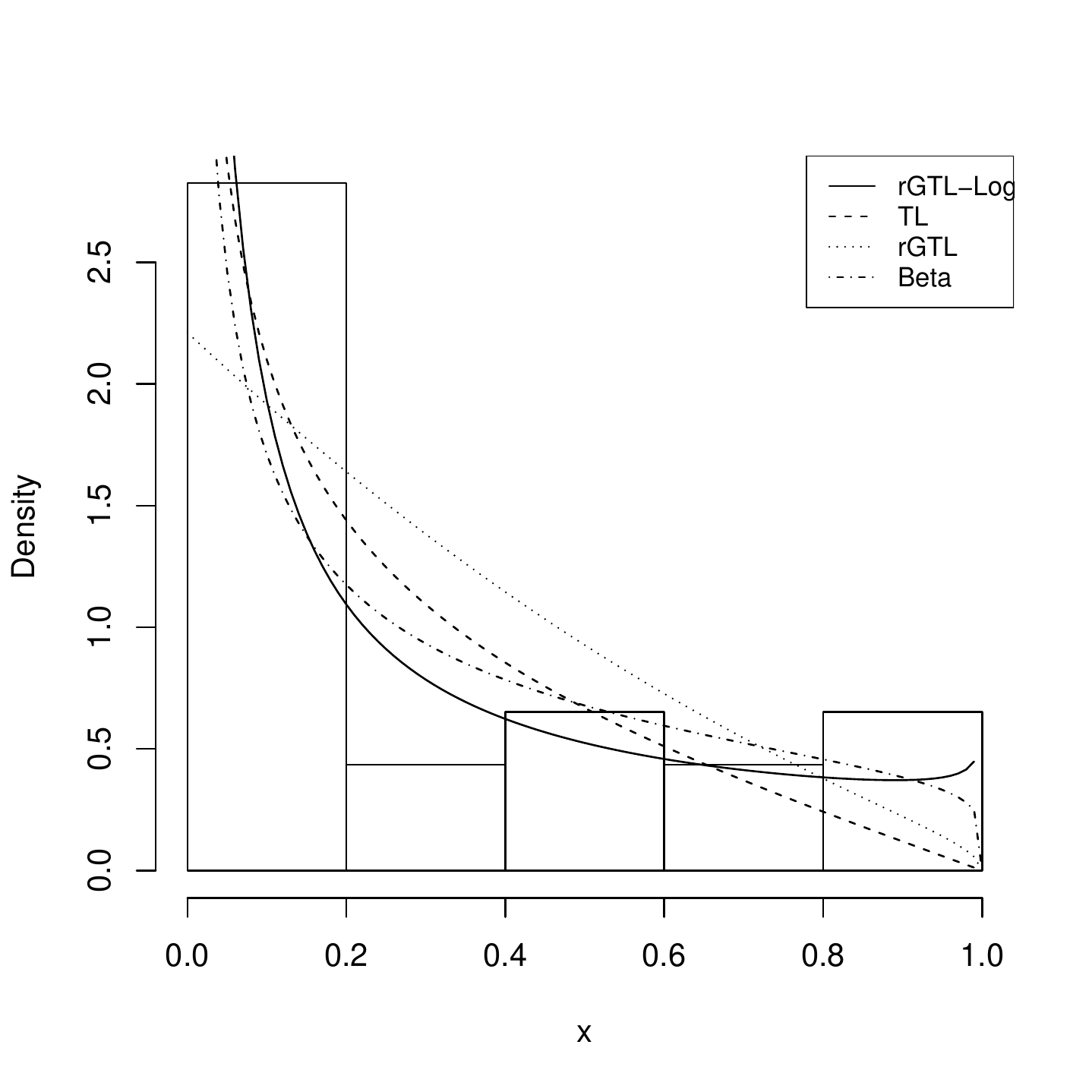} \includegraphics[width=0.48\textwidth]{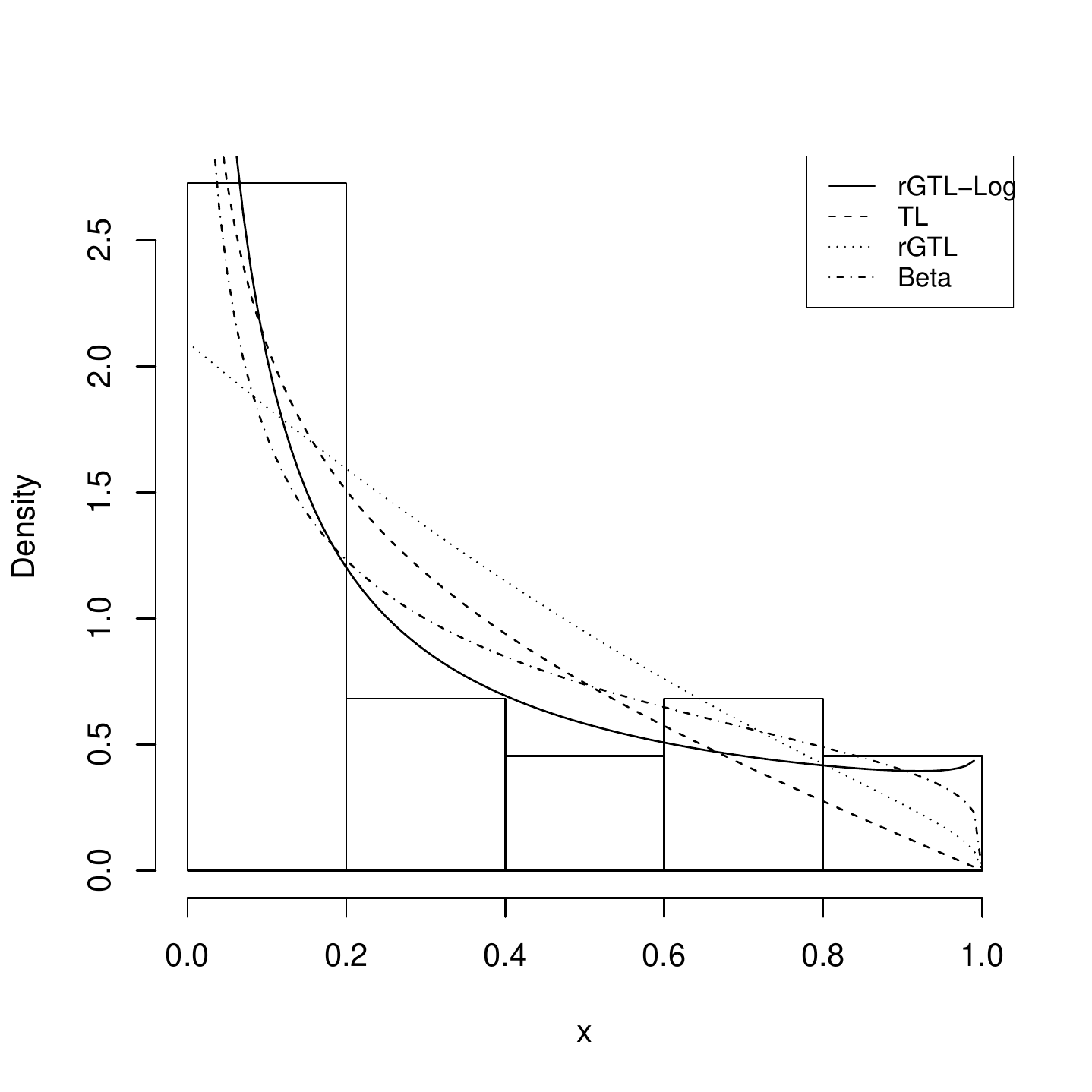}
  \caption{Empirical and fitted density functions for SC16 (left panel) and P3 (rigth panel) algorithms data.}\label{ecdfTL}
\label{fig:dens_capacity}
\end{figure}

Finally, in Fig. \ref{fig:dens_capacity} are shown, for the two algorithms, the fitted density for the \textit{rGTL-Log}, the \textit{TL}, the \textit{rGTL} and the Beta model. As it can be seen, also the plots confirm the superiority of the  \textit{rGTL-Log} model in both cases.
\\

\begin{table}[!ht]
\resizebox{\textwidth}{!}{
\begin{tabular}{*{7}c}
\hline
           &   rGTL-Log &   rGTL-Geo &   rGTL-Poi &       rGTL &         TL &       Beta \\
           &            &            &            &            &          &  \\
           &                                      \multicolumn{ 6}{c}{SC16} \\
\hline
$\hat{\alpha}$& 1.3980 (0.687) & 0.8856 (0.621) & 0.6184 (0.511) & 0.5444 (0.431) &            & a=0.4869 (0.121) \\
$\hat{\nu}$ & 0.8665 (0.455) & 0.5578 (0.576) & 1.0414 (0.544) & 1.5194 (0.518) & 0.5943 (0.1239) & b=1.1679 (0.358) \\
$\hat{\theta}$ & 0.9920 (0.014) & 0.9055 (0.119) & 2.1089 (1.311) &            &            &            \\
AIC &   -16.7599 &   -12.6145 &    -8.1251 &    -8.0792 &   -14.2302 &   -15.2149 \\
KS (p-value) & 0.1071 (0.9544) & 0.148 (0.6952) & 0.2376 (0.1491) & 0.3287 (0.0139) & 0.1690 (0.5272) & 0.1836 (0.4202) \\
\hline
           &            &            &            &            &            &            \\
           &                                                     \multicolumn{ 6}{c}{P3} \\
\hline
$\hat{\alpha}$ & 1.3275 (0.777) & 0.9098 (0.650) & 0.6455 (0.550) & 0.5573 (0.465) &            & a=0.5539 (0.142) \\
$\hat{\nu}$ & 0.9141 (0.475) & 0.6557 (0.583) & 1.0148 (0.562) & 1.4533 (0.523) & 0.6778 (0.145) & b=1.2198 (0.376) \\
$\hat{\theta}$ & 0.9821 (0.031) & 0.8611 (0.160) & 1.9458 (1.402) &            &            &            \\
AIC &   -10.6097 &    -8.2475 &    -5.3616 &     -6.342 &    -8.9965 &    -9.5638 \\
KS (p-value) & 0.1345 (0.8212) & 0.1432 (0.758) & 0.2383 (0.1642) & 0.3395 (0.0099) &  0.1848 (0.4400) & 0.2002 (0.3413) \\
\hline
\end{tabular}
}
\caption{\textit{ML} estimates of the parameters, AIC values and Kolmogorov-Smirnov test results for the first example dataset.}
\label{stime}
\end{table}

\begin{figure}
 \centering
 \includegraphics[width=0.48\textwidth]{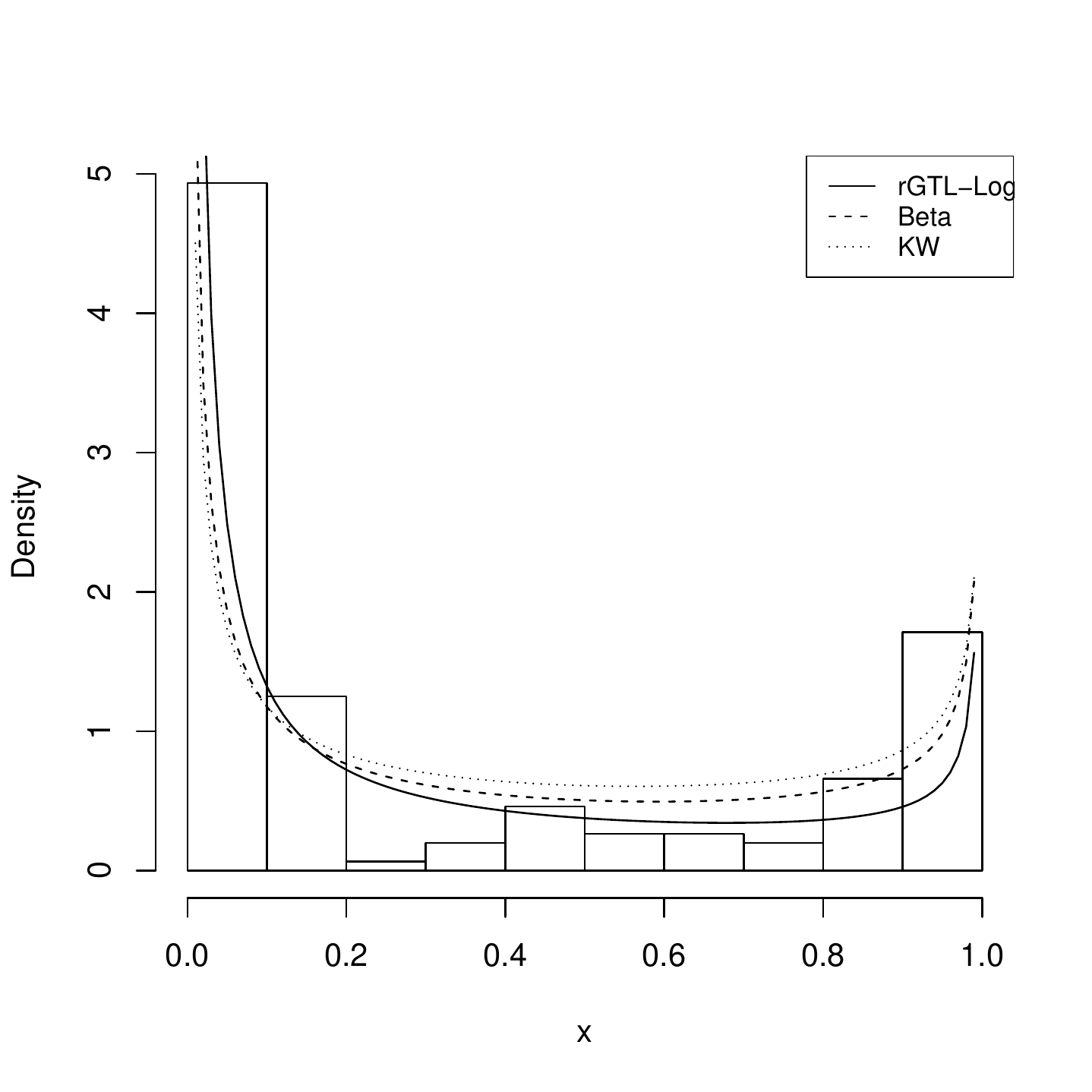} \includegraphics[width=0.48\textwidth]{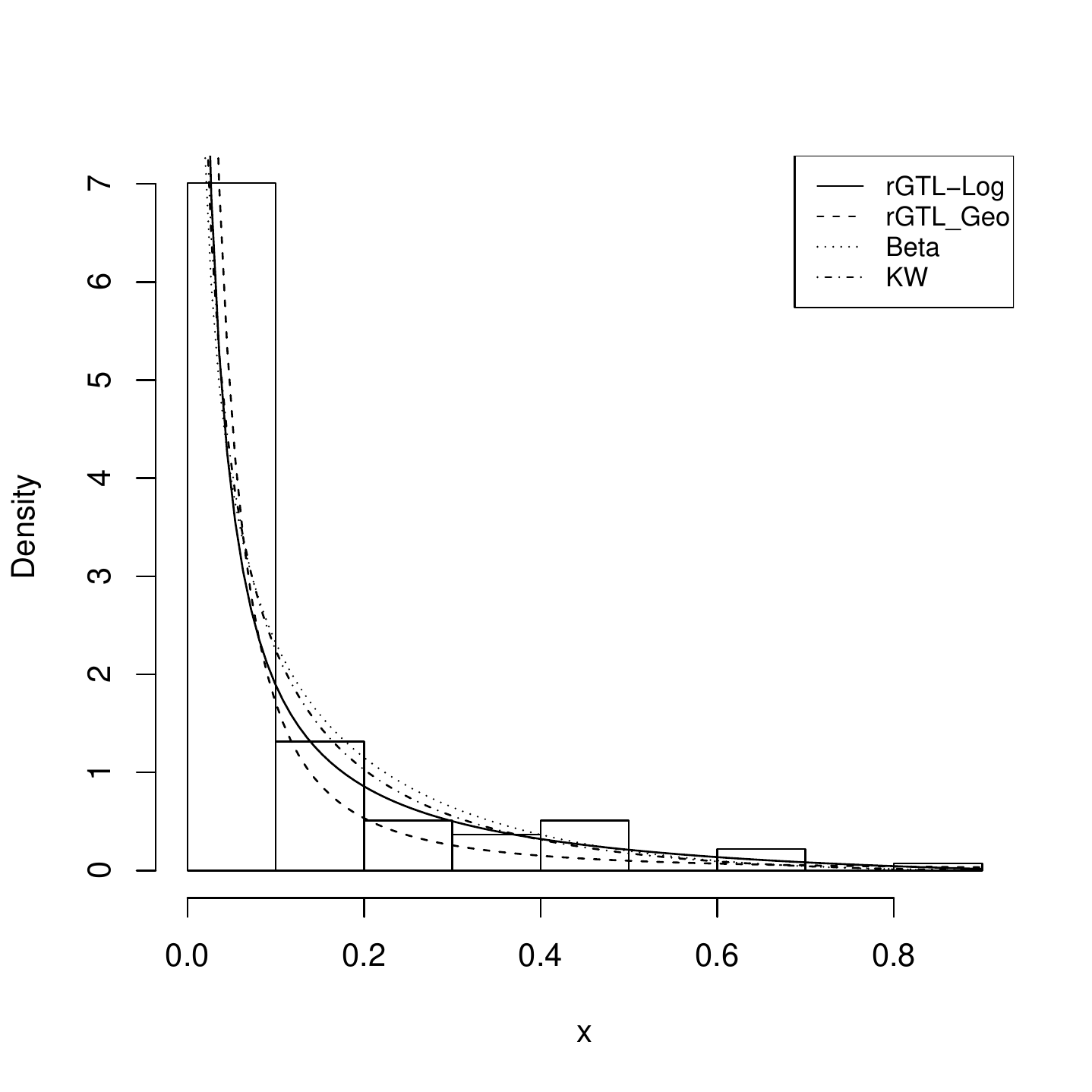}\hspace{0.2 cm}
  \caption{Empirical and fitted density functions for the proportions of muslim population and atheists.}
\label{fig:dens_ex2}
\end{figure}

\textit{Example 2.}
In this example, we consider the proportions of muslim population in 152 countries and the proportion of atheists of 137 countries. These datasets have been considered by \citet{Silva14} with the aim to select the best model between the Beta and the Kumaraswamy. Along these two models, we also consider \textit{rGTL-Log}, \textit{rGTL-Geo} and \textit{rGTL-Poi} models. The  \textit{ML} estimates of the parameters, with the corresponding standard errors, and the values for AIC  are reported in Table \ref{stime2}. In both cases, the \textit{rGTL-Log} model appears to be the best model, as suggested by the lowest value for the AIC. We note that, for the Atheism dataset, also the \textit{rGTL-Geo} seems to have a better performance than the  Beta and Kumaraswamy distributions.
In Fig. \ref{fig:dens_ex2} the fitted densities for the considered models are shown.

\begin{table}[!ht]
\resizebox{\textwidth}{!}{
\begin{tabular}{*{6}c}
\hline
           &   rGTL-Log &   rGTL-Geo &   rGTL-Poi &       Beta &         KW \\
           &            &            &            &            &            \\
           &                       \multicolumn{ 4}{c}{Muslim} &            \\
\hline
$\hat{\alpha}$ & 1.3997 (0.341) & 0.6521 (0.178) & 0.4573 (0.120) & a=0.2976 (0.028) & a=.2715 (0.033) \\
 $\hat{\nu}$ & 0.2624 (0.063) & 0.1226 (0.100) & 0.5530 (0.086) & b= 0.5159 (0.058) & b=0.5906 (0.057) \\
$\hat{\theta}$ & 0.9997 (3e-04) & 0.9796 (0.018) & 2.5828 (0.448) &            &            \\
       AIC &   -250.545 &   -150.499 &    -71.596 &   -232.908 &   -225.667 \\
           &            &            &            &            &            \\
           &                      \multicolumn{ 4}{c}{Atheism} &            \\
\hline
$\hat{\alpha}$ & 0.8411 (0.744) & 0.9952 (0.608) & 0.5353 (0.283) & a=0.4368 (0.043) & a=0.5091 (0.042) \\
 $\hat{\nu}$ & 2.3502 (1.242) & 0.9730 (0.758) & 3.0655 (0.782) & b= 3.6347 (0.538) & b=3.0914 (0.412) \\
$\hat{\theta}$ & 0.9900 (0.004) & 0.9746 (0.021) & 3.3155 (0.580) &            &            \\
       AIC &   -449.703 &   -438.989 &   -381.875 &   -407.951 &   -417.785 \\
\hline
\end{tabular}
}
\caption{\textit{ML} estimates of the parameters and AIC values for the second example dataset.}
\label{stime2}
\end{table}

\newpage

\section{Conclusion}
In many fields of applied science, the observations take values only in a limited range, it is the case, for example, of percentages, proportions and fractions. To model this type of data, the statistical literature offers very few alternatives, mainly the Beta distribution and only recently some authors have recovered the Topp-Leone distribution, proposed in 1955, and the Kumuraswamy distribution, introduced in literature in 1980. Certainly, the lack in the literature of distributions with bounded support contrasts with the huge presence of distributions with unbounded support. With the aim to reduce this gap, in this paper we have proposed a new class of distribution functions with limited support, namely \textit{rGTL-PS}, obtained by compounding the Power Series
distributions and the reflected Generalized Topp-Leone distribution. The proposed class includes, as special cases, some new distributions
with limited support such as the \textit{rGTL-Logarithmic}, \textit{the rGTL-Geometric}, the \textit{rGTL-Poisson} and \textit{rGTL-Binomial}.

Like the Beta distribution, the shape of the \textit{rGTL-PS} density function can be constant, increasing, decreasing, unimodal and uniantimodal depending on the values of its parameters. Unlike the Beta distribution, the hazard function of \textit{rGTL-PS} is much more flexible since it can be increasing, bathtub and N-shape. Moreover, the main advantage with respect to the Beta distribution is represented by the fact that the proposed model presents a distribution function in a closed form and the quantiles can be easly obtained. Finally, applications to some real data sets highlight the potential of the proposed model.

\newpage

\section{Appendix}
The partial derivatives of $g(y_i; \alpha, \nu)$ and $G(y_i; \alpha, \nu)$ with respect to $\alpha$ and $\nu$ are:
\begin{eqnarray*}
\dot{g}_{\alpha}(y_i; \alpha, \nu)= g(y_i; \alpha, \nu) \left\{  \frac{y_i (\nu-1)} {[\alpha-(\alpha-1)(1-y_i)]} + \frac{2y_i -1}{[\alpha-2(\alpha-1)(1-y_i)]}  \right\}
\end{eqnarray*}

\begin{eqnarray*}
\dot{G}_{\alpha}(y_i; \alpha, \nu)= -\nu y_i (1-y_i)^{\nu} [\alpha-(\alpha-1)(1-y_i)]^{\nu-1}
\end{eqnarray*}

\begin{eqnarray*}
\dot{g}_{\nu}(y_i; \alpha, \nu)= g(y_i; \alpha, \nu) \left\{  \frac{1}{\nu} + \ln(1-y_i) + \ln[\alpha-(\alpha-1)(1-y_i)]  \right\}
\end{eqnarray*}

\begin{eqnarray*}
\dot{G}_{\nu}(y_i; \alpha, \nu)= -\left\{(1-y_i) [\alpha-(\alpha-1)(1-y_i)] \right\} ^{\nu} \ln \left\{(1-y_i) [\alpha-(\alpha-1)(1-y_i)] \right\}
\end{eqnarray*}

\begin{eqnarray*}
\ddot{g}_{\alpha \alpha}(y_i; \alpha, \nu) = \left[\dot{g}_{\alpha}(y_i; \alpha, \nu)\right]^2 - g(y_i; \alpha, \nu) \frac{y_{i}^{2} (\nu-1)}{ [\alpha-2(\alpha-1)(1-y_i)] ^{2} }
\end{eqnarray*}

\begin{eqnarray*}
\ddot{G}_{\alpha \alpha}(y_i; \alpha, \nu)= -\nu (\nu-1) y_{i}^{2} (1-y_i)^{\nu} [\alpha-(\alpha-1)(1-y_i)]^{\nu-2}
\end{eqnarray*}

\begin{eqnarray*}
\ddot{g}_{\alpha \nu}(y_i; \alpha, \nu)= \dot{g}(y_i; \alpha, \nu) \left\{ \frac{y_i (\nu-1)} {[\alpha-(\alpha-1)(1-y_i)]} + \frac{2y_i -1}{[\alpha-2(\alpha-1)(1-y_i)]} \right\} + \frac{y_{i} g(y_i; \alpha, \nu)}{ [\alpha-(\alpha-1)(1-y_i)] }
\end{eqnarray*}

\begin{eqnarray*}
\ddot{G}_{\alpha \nu}(y_i; \alpha, \nu)= \dot{G}_{\alpha}(y_i; \alpha, \nu)   \left\{  \frac{1}{\nu} + \ln(1-y_i) + \ln[\alpha-(\alpha-1)(1-y_i)]   \right\}
\end{eqnarray*}

\begin{eqnarray*}
\ddot{g}_{\nu \nu}(y_i; \alpha, \nu)= \dot{g}_{\nu}(y_i; \alpha, \nu) \left\{  \frac{1}{\nu} + \ln(1-y_i) + \ln[\alpha-(\alpha-1)(1-y_i)]  \right\} -
\frac{g(y_i; \alpha, \nu)}{\nu^2}
\end{eqnarray*}

\begin{eqnarray*}
\ddot{G}_{\nu \nu}(y_i; \alpha, \nu)= \dot{G}_{\nu}(y_i; \alpha, \nu) \ln \left\{(1-y_i) [\alpha-(\alpha-1)(1-y_i)] \right\}.
\end{eqnarray*}

Putting $R_{1}(i)=\frac{ A^{'''} \left\{\theta[1-G(y_i; \alpha, \nu)]\right\} A^{'} \left\{\theta[1-G(y_i; \alpha, \nu)]\right\}-\left( A^{''} \left\{\theta[1-G(y_i; \alpha, \nu)]\right\} \right) ^2  }  { \left(A^{'} \left\{\theta[1-G(y_i; \alpha, \nu)]\right\}\right)^2 }$ and $R_{2}(i)=\frac{  A^{''} \left\{\theta[1-G(y_i; \alpha, \nu)]\right\}  }    { A^{'} \left\{\theta[1-G(y_i; \alpha, \nu)]\right\}}$, the elements of the observed information matrix $\bm {J}(\bm{\eta})$ are given by

\begin{eqnarray*}
J_{\alpha \alpha}=-\frac{\partial^2 \ell(\bm{\eta};\bm{y}) } {\partial \alpha ^2} &=& -\sum^{n}_{i=1} \frac{ \ddot{g}_{\alpha \alpha}(y_i; \alpha, \nu) g(y_i; \alpha, \nu) -[\dot{g}_{\alpha}(y_i; \alpha, \nu)]^2}{[g(y_i; \alpha, \nu)]^2} - \\
&& \theta^{2} \sum^{n}_{i=1} R_{1}(i) \left[\dot{G}_{\alpha}(y_i; \alpha, \nu)\right]^2 +
 \theta \sum^{n}_{i=1} R_{2}(i) \ddot{G}_{\alpha \alpha}(y_i; \alpha, \nu)
\end{eqnarray*}

\begin{eqnarray*}
J_{\alpha \nu}=-\frac{\partial^2 \ell(\bm{\eta};\bm{y}) } {\partial \alpha \partial \nu} &=& -\sum^{n}_{i=1} \frac{ \ddot{g}_{\alpha \nu}(y_i; \alpha, \nu) g(y_i; \alpha, \nu) -\dot{g}_{\alpha}(y_i; \alpha, \nu) \dot{g}_{\nu}(y_i; \alpha, \nu)}{[g(y_i; \alpha, \nu)]^2} - \\
&& \theta^{2} \sum^{n}_{i=1} R_{1}(i) \dot{G}_{\alpha}(y_i; \alpha, \nu) \dot{G}_{\nu}(y_i; \alpha, \nu) +
\theta  \sum^{n}_{i=1}  R_{2}(i) \ddot{G}_{\alpha \nu}(y_i; \alpha, \nu)
\end{eqnarray*}

\begin{eqnarray*}
J_{\alpha \theta}=-\frac{\partial^2 \ell(\bm{\eta};\bm{y}) } {\partial \alpha \partial \theta} &=&
  \sum^{n}_{i=1} R_{2}(i) + \theta \sum^{n}_{i=1} R_{1}(i) \dot{G}_{\alpha}(y_i; \alpha, \nu) \left[1-G(y_i; \alpha, \nu)\right]
\end{eqnarray*}

\begin{eqnarray*}
J_{\nu \nu}=-\frac{\partial^2 \ell(\bm{\eta};\bm{y}) } {\partial \nu^2} &=& -\sum^{n}_{i=1} \frac{ \ddot{g}_{\alpha \alpha}(y_i; \alpha, \nu) g(y_i; \alpha, \nu) -[\dot{g}_{\alpha}(y_i; \alpha, \nu)]^2}{[g(y_i; \alpha, \nu)]^2} - \\
&& \theta^{2} \sum^{n}_{i=1} R_{1}(i) \left[\dot{G}_{\nu}(y_i; \alpha, \nu)\right]^2 +
 \theta \sum^{n}_{i=1} R_{2}(i) \ddot{G}_{\nu \nu}(y_i; \alpha, \nu)
\end{eqnarray*}

\begin{eqnarray*}
J_{\nu \theta}=-\frac{\partial^2 \ell(\bm{\eta};\bm{y}) } {\partial \nu \partial \theta} &=&
 \sum^{n}_{i=1} R_{2}(i) + \theta \sum^{n}_{i=1} R_{1}(i) \dot{G}_{\nu}(y_i; \alpha, \nu) \left[1-G(y_i; \alpha, \nu)\right]
\end{eqnarray*}

\begin{eqnarray*}
J_{\theta \theta}=-\frac{\partial^2 \ell(\bm{\eta};\bm{y}) } {\partial \theta^2}= \frac{n}{\theta} + n\frac{A^{''}(\theta)A(\theta) - \left[A^{'}(\theta)\right]^2 }{\left[A(\theta)\right]^2}-\sum^{n}_{i=1} R_{1}(i) \left[1-G(y_i; \alpha, \nu)\right]^2.
\end{eqnarray*}

\end{document}